\documentclass[aps,reprint,showpacs,amsmath,amssymb,prl,superscriptaddress]{revtex4-1}

\usepackage[utf8]{inputenc}
\usepackage{latexsym}
\usepackage{color}
\usepackage{graphicx}
\usepackage{epsfig}
\usepackage{amsopn}
\usepackage{amsfonts}
\usepackage{amssymb}
\usepackage[hidelinks]{hyperref}
\usepackage{ wasysym }

\usepackage{subfigure}
\usepackage{braket}
\usepackage{epstopdf}
\usepackage{braket}
\usepackage{nicefrac}
\usepackage{amsmath}

\begin{document}

\title{A single-atom heat engine}

\author{Johannes Ro{\ss}nagel}
\email{j.rossnagel@uni-mainz.de}
\affiliation{Quantum, Institut f\"ur Physik, Universit\"at Mainz, D-55128 Mainz, Germany}

\author{Samuel Thomas Dawkins}
\affiliation{Quantum, Institut f\"ur Physik, Universit\"at Mainz, D-55128 Mainz, Germany}

\author{Karl Nicolas Tolazzi}
\affiliation{Quantum, Institut f\"ur Physik, Universit\"at Mainz, D-55128 Mainz, Germany}

\author{Obinna Abah}
\affiliation{Department of Physics, Friedrich-Alexander Universit\"at Erlangen-N\"urnberg, D-91058 Erlangen, Germany}

\author{Eric Lutz}
\affiliation{Department of Physics, Friedrich-Alexander Universit\"at Erlangen-N\"urnberg, D-91058 Erlangen, Germany}

\author{Ferdinand Schmidt-Kaler}
\affiliation{Quantum, Institut f\"ur Physik, Universit\"at Mainz, D-55128 Mainz, Germany}

\author{Kilian Singer}
\affiliation{Quantum, Institut f\"ur Physik, Universit\"at Mainz, D-55128 Mainz, Germany}
\affiliation{Experimentalphysik I, Universit\"at Kassel, Heinrich-Plett-Str. 40, D-34132 Kassel, Germany}

\date{\today}

\begin{abstract}
We report the experimental realization of a single-atom heat engine. An ion is confined in a linear Paul trap with tapered geometry and driven thermally by coupling it alternately to hot and cold reservoirs. The output power  of the engine is used to drive a harmonic oscillation. From direct measurements of the ion dynamics, we determine the  thermodynamic cycles for various temperature differences of the reservoirs.
We use these cycles to evaluate  power $P$ and efficiency $\eta$ of the engine, obtaining up to $P=342\,$yJ and $\eta=2.8 \,\permil$, consistent with analytical estimations. Our results demonstrate that thermal machines can be reduced to the ultimate limit of single atoms.
\end{abstract}

\maketitle


Heat engines have played a central role in our modern society since the industrial revolution. 
Converting thermal energy into mechanical work, they are ubiquitously employed  to generate motion, from cars to airplanes~\cite{cen01}. The working fluid of  a macroscopic engine typically contains  of the order of $10^{24}$ particles. In the last decade, dramatic experimental progress has lead to the miniaturization of thermal machines down to the microscale, using microelectromechanical~\cite{wha03}, piezoresistive~\cite{ste11} and cold atom~\cite{bra13} systems, as well as  single colloidal particles~\cite{bli12,mar14} and single molecules~\cite{hug02}. In his 1959 talk
``There is plenty of room at the bottom", Richard Feynman already envisioned tiny motors working at the atomic level~\cite{fey60}. However, to date no such device has been built. 


Here we report the realization of a single-atom heat engine whose working agent is an ion, held within a modified linear Paul trap.
 We use laser cooling and electric field noise to engineer cold and hot reservoirs.  
We further employ fast thermometry methods to determine the temperature of the ion~\cite{ros15}. 
The thermodynamic cycle of the engine
 is established for various temperature differences of the reservoirs, from which we deduce work and heat, and thus power output and efficiency.
We additionally show that the work produced by the engine can be effectively stored and used to drive an oscillator against friction. 
Our device demonstrates the working principles of a thermodynamic heat engine with a working agent reduced  to the ultimate single particle limit, thus fulfilling Feynman's dream.

Trapped ions offer an exceptional degree of preparation, control and measurement of their parameters, allowing for ground state cooling~\cite{lei03} and coupling to engineered reservoirs~\cite{mya00}. Owing to their unique properties, they have recently become invaluable tools for the investigation of quantum thermodynamics~\cite{hub08,an15,ulm13,aba12,ros14,ber13}.
They additionally provide an ideal setup to operate and characterize a single particle heat engine.

\begin{figure*}%
\centering
\includegraphics[width=0.9\textwidth]{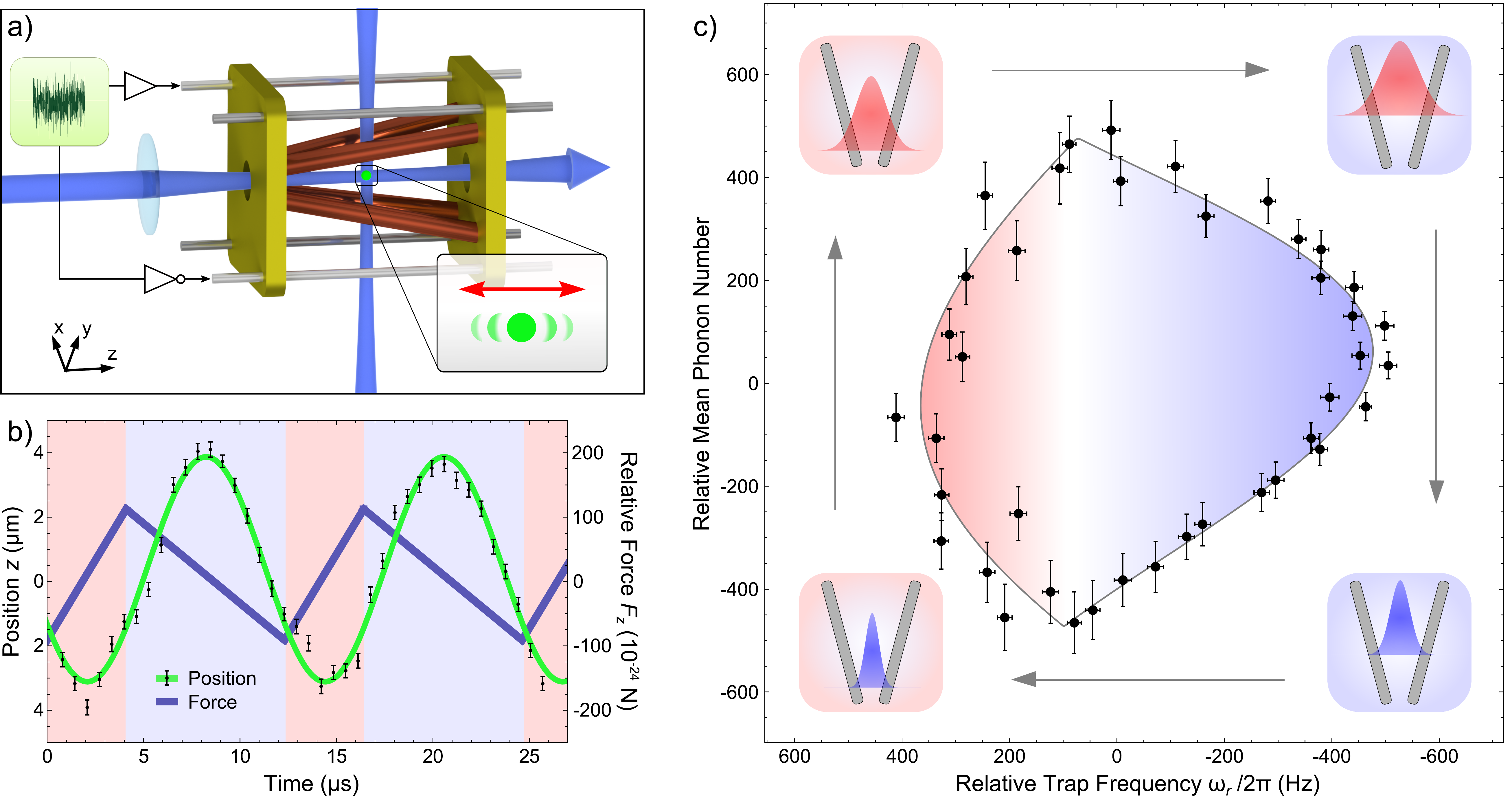}
\caption{
(a) Heat engine setup composed of a single trapped ion (green), lasers for cooling, damping and observation of the ion (blue), radio-frequency electrodes in funnel geometry (red), end-caps (gold) and outer electrodes (gray). The position of the ion is imaged on an ICCD camera. Opposing voltage noise waveforms are additionally supplied to the outer electrodes, in order to generate electric field noise without affecting the trap frequencies. 
(b) Position of the ion (black) determined from the average of more than $200\,000$ camera images at each time step. The error bars result from the uncertainty of Gaussian fits to the recorded fluorescence images.
The measured positions are described by a sinusoidal fit (green line). 
Background colors indicate the periodic interaction with the hot (red) and cold (blue) reservoirs which give rise to a periodic driving force (blue line) according to Eq.~(\ref{eq:zdependentconfinement}), shown relative to its mean value of $5.03\times 10^{-21}$. 
(c) Thermodynamic cycle of the engine for one radial direction: trap frequencies in the radial direction $\omega_r$ are deduced directly from the measured $z$-positions. The temperature $T$ of the radial state of motion and thus the corresponding mean phonon number $\bar{n}_r$ is determined from separate measurements (see text and Fig.~\ref{fig2}). The values of $\omega_r$ and $\bar{n}_r$ are given with respect to the center of the cycle at $\omega_{0r}/2\pi=447.9(2)\,$kHz and $\bar{n}_{0r}=26160(445)$. The shaded area enclosed by the cycle reflects the work performed by the engine, where red and blue colors indicate heating and cooling periods, respectively. The black line is the calculated trajectory of the cycle, see text. The pictograms in the corners illustrate the different strokes of an idealized cycle.}
\label{fig1}
\end{figure*}

In our experiment, a single $^{40}\textrm{Ca}^+$ ion is trapped in a linear Paul trap with a funnel-shaped electrode geometry, as shown in Fig.~\ref{fig1}a~\cite{aba12}. The electrodes are driven symmetrically at a radio-frequency voltage of $830\,\textrm{V}_\textrm{pp}$ at 21\,MHz, resulting in a tapered harmonic pseudo-potential~\cite{lei03} of the form
$U=({m}/{2})\sum_i{\omega_{i}^2 i^2}$,  
where $m$ is the atomic mass and $i\in\{x,y\}$ denote the trap axes as seen in  Fig.~\ref{fig1}a. 
The axial confinement is realized with constant voltages on the two end-cap electrodes, resulting in a trap frequency of $\omega_{z}/2\pi=81\,$kHz.
The trap angle $\theta=10^\circ$ and the radial extent of the trap $r_0=1.1\,$mm at $z=0$ characterize the geometry of the funnel. The resulting  radial trap frequencies $\omega_{x,y}$ decrease in the axial $z$-direction as
\begin{equation}
\omega_{x,y}=\frac{\omega_{0x,0y}}{(1 + z\,{\tan{\theta}}/{r_0})^2}.
\label{eq:zdependentconfinement}
\end{equation}
The eigenfrequencies in the radial directions  at the trap minimum $z=0$ are $\omega_{0x}/2\pi=447\,$kHz and $\omega_{0y}/2\pi=450\,$kHz, with the degeneracy lifted, but sufficiently close to permit the approximation of cylindrical symmetry with $r^2=x^2+y^2$ and a mean radial trap frequency $\omega_{r}$. 
 An additional set of outer electrodes is employed to 
compensate for stray fields. The trapped ion is cooled by a laser beam at $397\,$nm, which is red-detuned to the internal electronic $S_{1/2}-P_{1/2}$ transition~\cite{lei03}, and the resulting fluorescence is recorded by a rapidly-gated intensified charge-coupled device (ICCD) camera.

The heating and cooling of the ion is designed such that the ion thermalizes as if in contact with a thermal reservoir.
A cold bath interaction is realized by exposing the ion to a laser cooling beam, leading to an equilibrium temperature of $T_\textrm{C}=3.4\,$mK~\cite{ros15,met99}.
A hot reservoir interaction with finite temperature $T_\textrm{H}$ is designed by  additionally   exposing the ion to white electric field noise. The interplay of photon scattering and noise leads to a thermal state  of the ion at temperature $T$ at any given moment \cite{ros15,cir92,cir94}.

In our setup,  heating and cooling act on the radial degrees of freedom. The resulting  time-averaged spatial distribution of the thermal state is of the form,
\begin{equation}
\xi_\textrm{r}(r,\phi,T)=\frac{1}{2\pi \sigma_\textrm{r}^2} \exp\left[{\frac{-(r-r_0)^2}{2 \sigma_\textrm{r}^2}}\right],
\label{eq:}
\end{equation}
with a temperature-dependent time-averaged width $\sigma_\textrm{r}=\sqrt{k_B T/m \omega_{r}^2}$, where $k_B$ is the Boltzmann constant.
Owing to the geometry  of  the funnel potential, the ion experiences  an average   force in  axial direction given by,
\begin{equation}
F_z(T)=-\int_0^\infty \xi_\text{r}(r,\phi,T) \frac{{d}U}{{dz}}\,{d\phi dr}.
\label{eq:force}
\end{equation}
The heat engine is  driven by alternately heating and cooling the ion in radial direction by switching the electric noise on and off; the cooling laser is always on.
Heating expands the width of the radial thermal state. As a result, the ion moves along the $z$-axis to a weaker radial confinement. We calculate a static displacement of $11\,$nm for the relative change of $F_z$ corresponding to Fig.~\ref{fig1}. 
During this first step of the engine, the axial potential energy of the ion increases and work is produced.
The second step occurs during exposure to the cold reservoir when the electric field noise is switched off.  Here the radial width $\sigma_\text{r}$ and the corresponding force $F_z$  decrease as the temperature is reduced, and the ion moves back to its initial position owing to the restoring force of the axial potential.
The combination of heating and cooling give rise to a closed thermodynamic cycle and leads to a periodic force $F_z(T)$ in the axial direction (see Fig. 1c).
When the cycle repeats itself at a rate which is close to the axial trap frequency, the engine effectively drives the harmonic oscillation. The work produced in each radial cycle is then transferred to the axial degree of freedom and stored in the amplitude of the oscillation. The essentially frictionless nature of the system leads to an ever-increasing oscillation. The axial motion thus plays a role similar to the flywheel of a mechanical engine.

In order to contain this oscillation, we provide adjustable damping by introducing an additional cooling laser in the axial direction. 
Steady state operation is reached when the work generated by the engine is balanced by the energy dissipated by the damping.
We measure the amplitude of the steady state oscillation of the ion in the $z$-direction by recording fluorescence images with the ICCD camera, using an exposure time of 700\,ns which is much shorter than the axial oscillation period. 
The camera is synchronized with the temperature modulation, allowing for the repetitive recording of images at particular phases of the oscillation.
Thus, the position of the ion can be determined precisely as a function of time  revealing the amplitude and phase of the resulting sinusoidal oscillation~\cite{nag98} (see Fig.~\ref{fig1}b). We verified that both are independent to an inversion of the noise signal. 
To facilitate the position measurement the damping laser is employed to reduce the incoherent, thermal motion in the axial direction, which originates from bath interactions.
In order to have efficient cooling whilst maintaining low damping of the oscillation, the laser is applied only around the turning points of the axial motion \cite{met99}. 


\begin{figure}%
\centering
\includegraphics[width=\columnwidth]{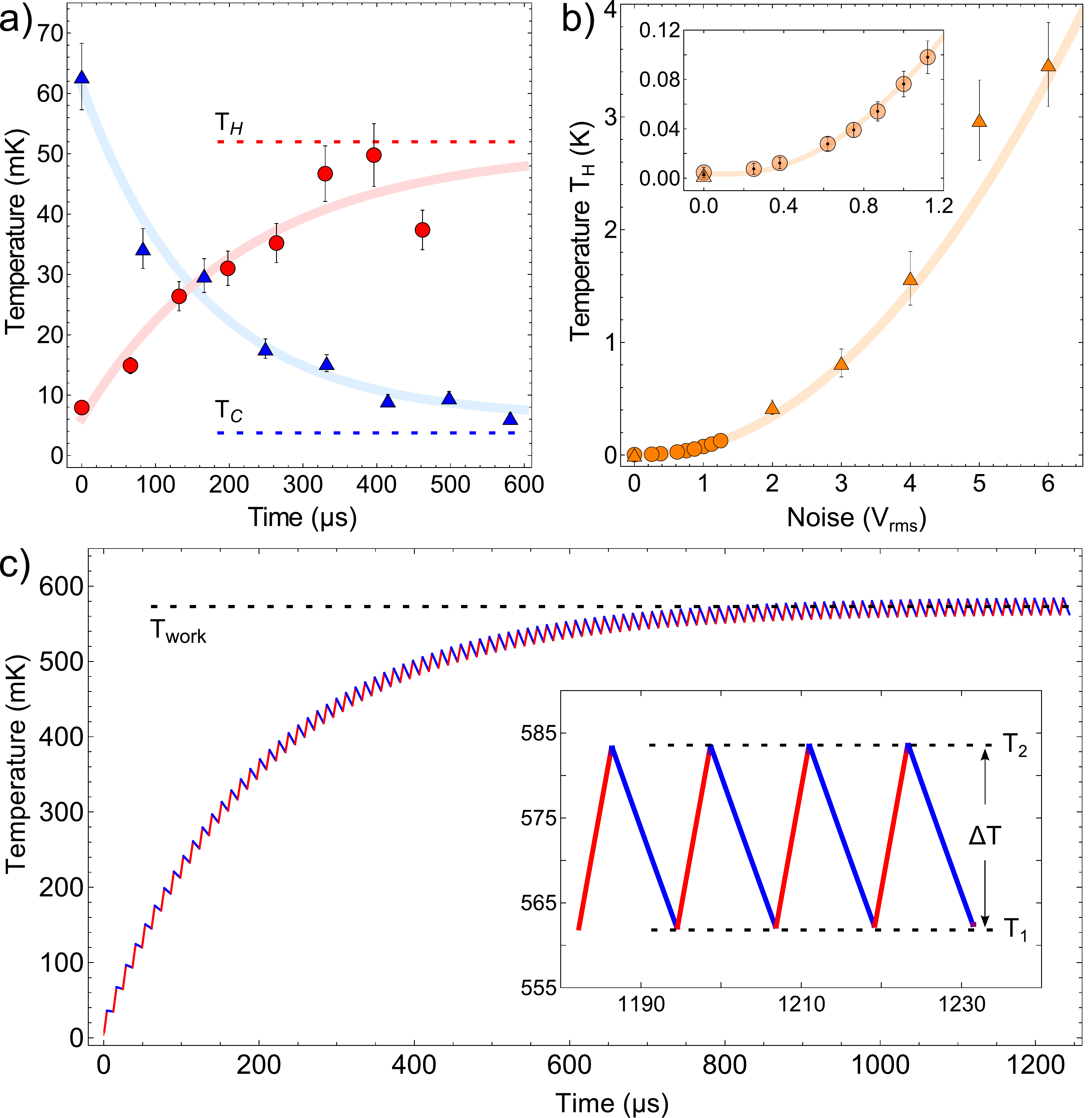}
\caption{Temperature dynamics of the engine. (a) Thermalization curves, derived from dark-state thermometry of the radial thermal state of the ion when heated to $T_{H}$ (red) and cooled to $T_C$ (blue). Individual errors result from fits to the dark-resonances as well as the uncertainty due to the laser-linewidths~\cite{ros15}. Individual fits reveal thermalization time constants for heating and cooling of $189(26)\,\mu$s and $216(13)\,\mu$s, respectively. (b) Dependence of the hot bath temperature $T_{H}$ on the root-mean-square of the electric noise applied to the trap electrodes, measured by dark-state thermometry (circles) \cite{ros15} and spatial thermometry (triangles) \cite{knu12}. (c) Simulated temperature of the ion as a function of time using the parameters for thermalization. Heating and cooling durations in each cycle are $4.1\,\mu$s and $8.2\,\mu$s, respectively. With an electric noise amplitude of $4\,$V, we calculate a working point temperature $T_\text{work}=568\,$mK and a temperature difference  $\Delta T=T_2-T_1=21.5\,$mK.
}%
\label{fig2}%
\end{figure}

The thermodynamic cycle of the single ion heat engine is presented in Fig.~\ref{fig1}b by plotting the mean phonon number $\bar{n}_\textrm{r}=k_B T/\hbar \omega_r$ of the thermal state of the ion in the radial direction as a function of the corresponding trap frequency $\omega_\textrm{r}$~\cite{lin03}. The radial trap frequencies $\omega_\textrm{r}(t)$ are obtained from the measured axial positions of the ion $z(t)$ in conjunction with calibration measurements of the tapered confinement described by Eq.~(\ref{eq:zdependentconfinement}). The temperature of the ion at any given moment of the cycle is deduced from the interplay of the heating and cooling rates, as illustrated in Fig.~\ref{fig2}. We determine these rates via a stroboscopic measurement of the thermal broadening of narrow resonances of dark-states caused by coherent population trapping~\cite{ros15,pet12} (Fig.~\ref{fig2}a).
The temperature of the hot reservoir $T_H$ can be adjusted via the applied electric field noise, and has been measured using dark-state thermometry, as well as spatial thermometry~\cite{knu12} for higher temperatures (Fig.~\ref{fig2}b). 
The heating and cooling processes are much slower than the internal dynamics of the ion and thus the cycle can be regarded as quasistatic with negligible losses due to irreversible processes~\cite{aga13}.
As the ion is in permanent contact with one of the reservoirs, the dynamics are similar to those of a Stirling engine~\cite{bli12,bra15}.


\begin{figure}%
\centering
\includegraphics[width=\columnwidth]{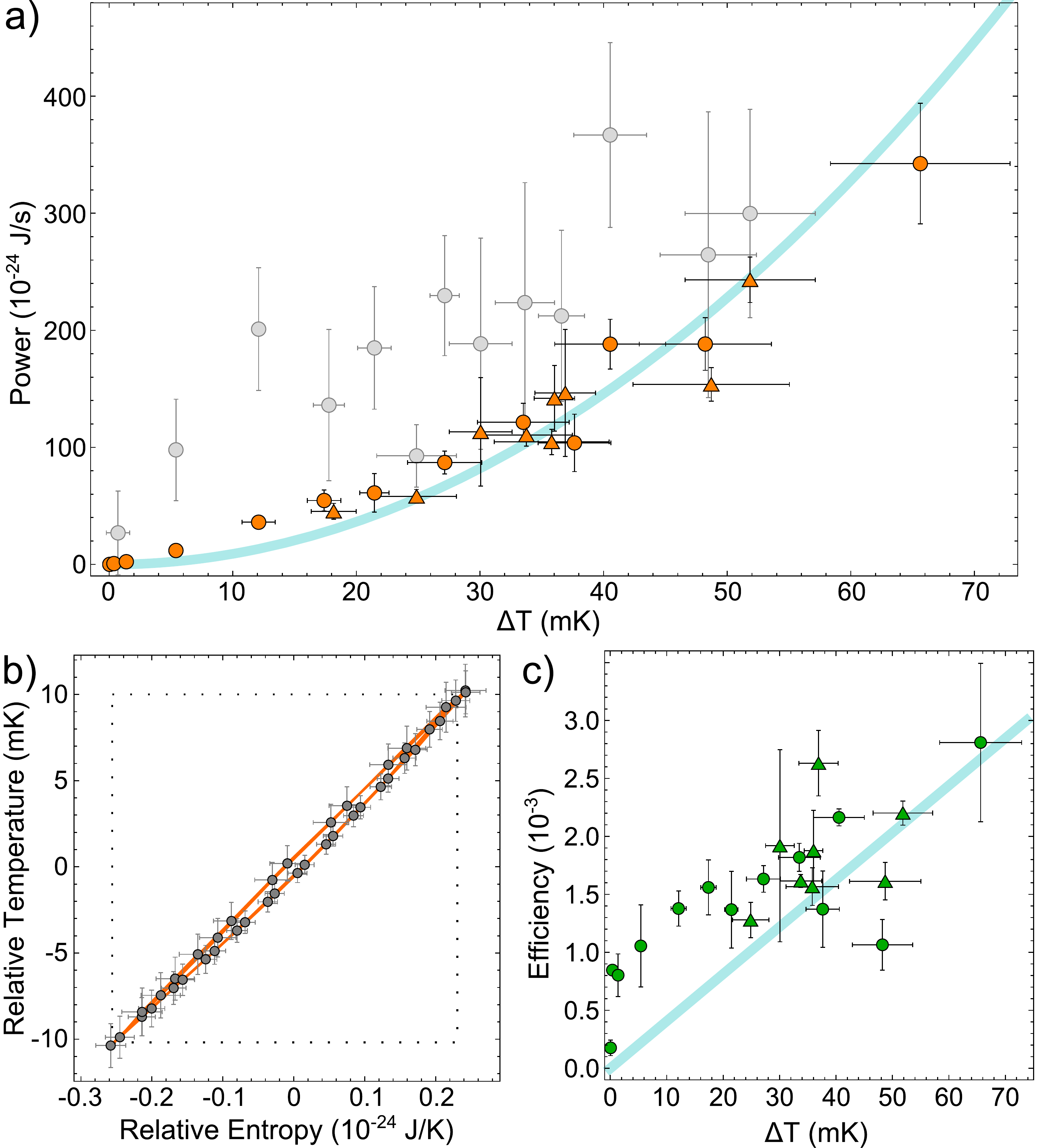}
\caption{(a) Power of the engine, calculated from the measured  cycles (orange) and from a direct analysis of the amplitude (gray).
The temperature differences $\Delta T$ were achieved by varying the electric field noise (circles) and the duty cycle of heating and cooling (triangles). The measured cycle data is consistent with the expected value from  analytical calculations (blue line). (b) The heat engine cycle, correpsonding to Fig.~\ref{fig1}c, measured as temperature relative to $568\,$mK versus  entropy relative to $179.6\times 10^{-24}\,$J/K. Both work and efficiency can be derived from integrating this cycle. The square (dotted line) represents the equivalent Carnot cycle for the full range of parameters and shows the case of the theoretical maximum efficiency. (c) Measured efficiency as a function of $\Delta T$, by varying the noise amplitude (circles) or the duty cycle (triangles), compared to the result of the analytical calculation.
}%
\label{fig3}%
\end{figure}

We derive the power output   during steady state operation in three independent ways.
We first determine the power  $P_\textrm{c}=W_\textrm{c}/t_\textrm{c}$, with cycle time  $t_\textrm{cyc}=2\pi/\omega_z$, by evaluating  the work as the area of the cycles for both radial directions,  $W_\textrm{c}=\hbar \oint  2\bar{n}_r d\omega_\textrm{r}$.
To assess the performance of the engine, we compute the power for various temperature differences $\Delta T=T_2-T_1$ between maximum temperature $T_2$ and minimum temperature $T_1$ in the cycle, as defined in Fig.~\ref{fig2}c. $\Delta T$ can be tuned by adjusting either the reservoir temperatures or the relative duration of the reservoir interaction, given by the duty cycle $d=t_H/t_\textrm{cyc}$ of the hot bath interaction time $t_H$ per cycle. 

We alternatively deduce the power directly from the measurement of the axial oscillation amplitudes $A_z$ of up to $15\,\mu$m. 
The driving power of a driven damped harmonic oscillator at steady state yields 
 $P_\textrm{o}=\gamma m \omega_z^2 A_z^2$ \cite{tay05}.  
The damping parameter $\gamma$ was determined separately via observation of the decay of the oscillation to be $\gamma=481\pm 141\,\textrm{s}^{-1}$. 
Both methods give  consistent values of the power,  in the range of $10^{-22}$ J/s, depending on $\Delta T$ (see Fig.~\ref{fig3}a). 
This represents a power-to-mass ratio of $1.5\,\textrm{kW}/\textrm{kg}$, comparable to that of a typical car engine.

We further calculate analytically the engine output power using the expression for work performed during a single cycle,
\begin{equation}
W_\text{a}=-\oint{F_z(t) \,\frac{{d}z(t)}{{d}t}{d}t}.
\label{eq:}
\end{equation}
The driving force $F_z$ is calculated from Eq.~(\ref{eq:force}) accounting for  the temperature variations of the ion determined as shown in Fig.~\ref{fig2}c, and neglecting the  weak $z$ dependence. The resulting motion $z(t)$ is derived assuming expressions of a resonantly driven damped harmonic oscillator.
Thus, for the output power we find $P_\textrm{a}=W_a/t_\textrm{cyc}= 
 {8 k_B^2 \sin^2(\pi  d)  \tan^2 \theta \Delta T^2}/ [{m \pi^3 \gamma \omega_z} ((d^2 - d) r_0)^2]={9.1 \times10^{-20} \Delta T^2}$. 
This analytical formula is plotted using experimental parameters in Fig.~\ref{fig3}a and is in agreement with the measured~$P_\textrm{c}$.


We further evaluate the efficiency of the engine, $\eta_\textrm{c}=W_\textrm{c}/Q_H$, from the measured data by determining the heat absorbed from the hot reservoir, $Q_H=\int_H T {d}S$, from the $TS$ diagram shown in Fig.~\ref{fig3}b. 
Here $S$ denotes the entropy of a thermal  harmonic oscillator, $S=k_B \left[ 1+\ln \left( k_B T/(\hbar \omega_r) \right) \right]$~\cite{aga13,lan80}. To this end, we transform the  measured data of the cycle in Fig.~2b  according to $\{\omega_r,\bar{n}_r\} \rightarrow \{S,T \}$. Employing the above analytical approach, we find $\eta_a = 4 k_B \sin^2(\pi  d)  \tan^2 \theta \Delta T/ [m \pi^3 \gamma \omega_z ((d^2 - d) r_0)^2]=0.041 \Delta T$.
The resulting efficiencies are displayed in Fig.~\ref{fig3}c, and reach values of up to $\eta_\textrm{c}= 0.28\,\%$, in agreement with the analytical expectation.
A comparison of this value with the corresponding efficiency at maximum power,  given by the Curzon-Ahlborn formula $\eta_\textrm{CA}= 1-\sqrt{T_1/T_2} {= 1.9\%}$~\cite{cur75}, reflects that the current trap parameters do not correspond to the optimal point~\cite{aba12}.
The performance of future single ion heat engines could be improved by redesigning the geometry of the trap to have  cycles with a higher range of frequencies $\omega_r$ (see Fig.~\ref{fig3}b). This  could be achieved by increasing either the angle of the taper or the absolute radial trap frequencies.


In this paper, we have demonstrated a first realization of a heat engine whose working agent is a single atom. The versatility of the trapped ion system allowed for the measurement of key features such as thermodynamic cycles, power output and efficiency. Ion trap technology features exquisite control, even down to the limit of single quanta, and permits coupling to designed reservoirs [18], state preparation and read-out. A future single ion heat engine operating close to the ground state of motion may reach the quantum regime and employ genuine quantum effects in thermodynamics \cite{scu03,dil09,pek15}.

 We thank Princeton Instruments for the loan of the ICCD camera and S. Deffner for fruitful discussions.
 We acknowledge the support of the German Research Foundation (grant ``Einzelionenw\"armekraftmaschine"), the Volkswagen Foundation (grant ``Atomic Nano-Assembler"), as well as the EU COST action MP1209 and  the
EU Collaborative Project TherMiQ (grant agreement 618074).

\end{document}